\documentstyle[12pt]{article}
\newcommand{\newsection}{ \setcounter{equation}{0} \section}
\def\appendix#1{
  \addtocounter{section}{1}
  \setcounter{equation}{0}
  \renewcommand{\thesection}{\Alph{section}}
  \section*{Appendix \thesection\protect\indent #1}
  \addcontentsline{toc}{section}{Appendix \thesection\ \ \ #1}
  }
\newcommand{\rf}[1]{(\ref{#1})}
\newcommand{\beq}{\begin{equation}}
\newcommand{\eeq}{\end{equation}}
\newcommand{\bea}{\begin{eqnarray}}
\newcommand{\eea}{\end{eqnarray}}
\newcommand{\tr}{\hbox{\rm Tr}\;}
\newcommand{\mgn}{{{\cal M}_{g,n}}}

\newcommand{\gn}{{g,n}}

\newcommand{\la}{\left\langle}
\newcommand{\ra}{\right\rangle}
\newcommand{\zet}{\hbox{\bf Z}}
\newcommand{\smallzet}{\mbox{\scriptsize\bf Z}}

\newcommand{\calt}{{\cal T}}
\newcommand{\calh}{{\cal H}}

\newcommand{\cald}{{\cal D}}

\newcommand{\x}{{\cal X}}
\newcommand{\comb}{{\,\mbox{\scriptsize dec}}}
\newcommand{\aut}{{\hbox{Aut}\;}}

\renewcommand{\d}{{\, d}}

\begin{document}
\topmargin 0pt
\oddsidemargin 5mm
\headheight 0pt
\headsep 0pt
\topskip 9mm

\addtolength{\baselineskip}{0.20\baselineskip}
\hfill hep-th/9402030

\hfill IFUM - 459/FT

\hfill January 1994
\begin{center}

\vspace{30pt}
{\large \bf
Moduli Spaces of Curves with Homology Chains\\
and $c=1$ Matrix Models
}

\vspace{30pt}
{\sl A.~S.~Cattaneo}

\vspace{10pt}
{\footnotesize
Dipartimento di Fisica, Universit\`a di Milano\\
and INFN, Sezione di Milano, \\
via Celoria 16, 20133 Milano, Italy\\
E.mail: {\it cattaneo@vaxmi.mi.infn.it}}

\vspace{20pt}

{\sl A.~Gamba}

\vspace{10pt}
{\footnotesize
Dipartimento di Matematica, Politecnico di Torino\\
viale Duca degli Abruzzi 24, 10129 Torino, Italy\\
and INFN, Sezione di Pavia, 27100 Pavia, Italy\\
E.mail: {\it gamba@polito.it}}

\vspace{20pt}

and

\vspace{20pt}

{\sl M.~Martellini}

\vspace{10pt}
{\footnotesize
Dipartimento di Fisica, Universit\`a di Milano,\\
via Celoria 16, 20133 Milano, Italy\\
and INFN, Sezione di Pavia, 27100 Pavia, Italy \\
E.mail: {\it martellini@vaxmi.mi.infn.it}}

\end{center}

\vspace{20pt}

\vfill

\begin{center}
{\bf Abstract}
\end{center}
We show
that introducing a periodic time coordinate in the models of
Penner-Kontsevich type generalizes
the corresponding constructions to the case of the
moduli space ${\cal S}_\gn^k$
of curves $C$ with homology chains $\gamma\in H_1(C,\zet_k)$.
We make a minimal extension of the resulting models
by adding a kinetic
term, and we get a new matrix model which realizes a simple
dynamics of $\zet_k$-chains on surfaces.
This gives a representation
of $c=1$ matter coupled
to two-dimensional quantum gravity
with the target space being a circle of finite radius, as
studied by Gross and Klebanov.

\vfill

\newpage

\newsection{Introduction}

The models of Penner-Kontsevich type
\cite{penner,kontsevich}
make use of a cellular
decomposition of the moduli space
$\mgn$
of algebraic curves which relies on the following fact~\cite{strebel}:
smooth algebraic curves
$C$
of genus
$g$
with
$n$
marked points
$x_1,\ldots,x_n$
and positive numbers
$p_1,\ldots,p_n$
are in 1-1 correspondence with ribbon graphs with metric~\footnote{
A ribbon graph with metric is a graph with
a cyclic ordering of the edges around each vertex and
with assigned positive lengths of the edges.
These graphs arise as set of separatrices of
the closed geodesics of certain minimal area metrics on
$C$
and carry all the information about the complex structure of the
curve.
}.
This combinatorial description greatly simplifies
the sum over the space of isomorphism classes of surfaces,
which becomes substituted by a sum over isomorphism classes of
graphs.

{}From
the point of view of two-dimensional quantum gravity
it would be interesting to get an
analogous description not only for curves, but also for curves carrying
additional structures.
As a matter of fact, the sum over surfaces corresponds to a
theory of pure two-dimensional quantum gravity, while
matter fields should be described by ``spin'' structures.
Three years ago, E.~Witten~\cite{witten}
has proposed a model of 2D-gravity interacting with matter fields
which involves a sum over the moduli space of curves with
$k$-th roots of the canonical bundle.
The model we discuss in this paper was inspired by Ref.~\cite{witten}
but differs from it on several aspects.
First of all, we consider here the moduli space of curves with
fixed $\zet_k$-chains, which
do not coincide with $k$-th
roots of the canonical bundle since the correspondence between the
two objects is non-canonical and depends on the choice of a reference
$k$-th root.
Secondly, we were not able at this stage to describe combinatorially
the Euler class which is an essential ingredient of Witten's paper.
On the other hand, we were able to give a natural combinatorial
description of $\zet_k$-chains
which fits very well in the Penner-Kontsevich picture.
We have shown that
it is possible to give a cellular decomposition
of the moduli space ${\cal S}_\gn^k$ of
couples $(C,\gamma)$ with
$\gamma\in H_1(C,\zet_k)$
and that there exist a matrix model realization generating
this cellular decomposition.
This is obtained by simply substituting
the usual hermitian matrix by a hermitian matrix field depending
on a time variable defined on a circle of finite radius $R$.
This
gives a general method for extending models of the
Penner-Kontsevich type to the moduli space of curves with spin
structures.
However, this is a non-interacting picture, since the new partition
function factorizes in a product of the old ones.
But now we are in a position to make a minimal modification
to these models by introducing a kinetic
$(\dot M(\phi))^2$
term.
The resulting model is a representation of $c=1$ matter coupled
to two-dimensional quantum gravity with the target space being the
circle of radius $R$.
This model has been studied by Gross and Klebanov~\cite{grosskleb}.
It presents different behaviours for
$R>R_{\mbox{\scriptsize c}}$ and $R<R_{\mbox{\scriptsize c}}$, and
a Kosterlitz-Thouless phase transition for $R=R_{\mbox{\scriptsize c}}=1$.
For $R\rightarrow+\infty$ the non-singlet states of the model decouple.
These states are identified with vortices on the world sheet, and
they become relevant for $R<R_{\mbox{\scriptsize c}}$.
In our construction the parameter $R$ interpolates between
the topologically trivial situation $R=+\infty$ and a new topological
phase which has yet to be investigated.
We conjecture that the model at the critical point
$R=R_{\mbox{\scriptsize c}}=1$
should describe some non-trivial topological characteristics of the
spaces ${\cal S}_\gn$.

\newsection{Combinatorial description of $H_1(C,\zet_k)$}

Let us introduce some notation.
With
$C_\gn$
we will denote a smooth algebraic curve of genus
$g$
with
$n$
punctures
$x_1,\ldots,x_n$
and
$n$
positive numbers
$p_1,\ldots,p_n$
associated to them.
$\x$ will be
the corresponding fat graph with metric, obtained from the
Strebel-Kontsevich construction.
Let
$X$
be the set of oriented edges of
$\x$,
$s_0$
be the permutation of
$X$
which clockwise exchanges edges with a common source,
$s_1$
the orientation-exchanging permutation, and
$s_2=s_0s_1$.
Finally we set
$X_j=X/s_j$,
$j=0,1,2$
and denote with
$[x]_j$
the equivalence class of
$x\in X$
in
$X_j$.

Let
$C_\gn'=C_\gn\backslash\{x_1,\ldots,x_n\}$
and consider the group
$H_1(C_\gn',\zet_k)$
of homology chains with coefficients in
$\zet_k$
(the cyclic group of order $k$).
We will show that
$H_1(C_\gn',\zet_k)$
is canonically isomorphic to the group
$G_\x$
of colorings
$c:X_1\rightarrow\zet_k$
such that
\bea
\label{staruno}
\sum_{x\in[y]_0}c(x) &\equiv& 0\quad (\hbox{mod}\, k)
\quad\hbox{for all}\; y\in X_1,   \\
\label{stardue}
\sum_{x\in[y]_1}c(x) &\equiv& 0\quad(\hbox{mod}\,k)
\quad\hbox{for all}\; y\in X_1.
\eea
As a matter of fact,
let us consider the application
$\phi_y:\x\rightarrow\x'$
which cancels the edge
$y\in X_1$
and identifies the corresponding vertices.
It is easily seen that
$\phi_y$
induces an {\em iso\/}morphism
$G_\x\rightarrow G_{\x'}$.
Therefore, we can reduce to the case of a graph
$\tilde\x$
with
$2g-1+n$
edges and only one vertex,
homotopically equivalent to
$\x$.
The edges
$x_j$,
$j=1,\ldots,2g-1+n$
are now closed cycles and generate
$H_1(\tilde\x,\zet_k)$.
We can therefore identify
$c(x_j)\in\zet_k$
with the coefficient of the cycle
$x_j$,
thus getting
$G_\x\simeq G_{\tilde\x}\simeq H_1(\tilde\x,\zet_k)\simeq H(\x,\zet_k)$.

Let us now define
${\cal S}_\gn^k$
as the moduli space of couples
\beq
(C_\gn,\gamma)
\label{starstar}
\eeq
where
$\gamma\in H_1(C'_\gn,\zet_k)$
is a
$\zet_k$-chain
on the smooth algebraic curve
$C'_\gn$
deprived of the $n$ marked points $x_1,\ldots,x_n$.
The datum~\rf{starstar}
can be combinatorially realized using the colored graphs described
in~\rf{staruno} and~\rf{stardue}.
These graphs provide a cell decomposition of
${\cal S}_\gn^{k,\comb}$
in the same way as the usual Strebel graphs provide a cell
decomposition of
${\cal M}_\gn^\comb$~\footnote{
The suffix ``dec'' (``decorated'') refers to the fact that each curve
carries the additional datum of $n$ positive numbers $p_1,\ldots,p_n$
associated to the punctures.
}.

\newsection{Vertices}

Let
$\{M_a\}_{a\in\smallzet_k}$
be a collection of matrices
$M_a\in gl(N)$.
Let us consider terms of the form
$\tr(M_{a_1}\ldots M_{a_n})$.
These terms are conveniently represented as colored
fat vertices
$v=[(a_1,\ldots,a_n)]\in V_n$,
where
\beq
V_n=\{(a_1,\ldots,a_n)\in(\zet_k)^n\}/\sim
\eeq
and
$\sim$
represents equivalence with respect to cyclic rotations of the
indices.
We have the formula
\beq
\tr\left(\sum_{a\in\smallzet_k}M_a\right)^n
=\sum_{v\in V_n}{n\over\#\aut v}\tr M_{a_1(v)}\ldots M_{a_n(v)}.
\label{dot}
\eeq
As a matter of fact,
\beq
\tr\left(\sum_{a\in\smallzet_k}M_a\right)^n
=\sum_{(a_.)\in(\smallzet_k)^n}\tr M_{a_1}\ldots M_{a_n}
\eeq
and the group
$\zet_k$
of cyclic rotations acts on the set of
$\tr M_{a_1}\ldots M_{a_n}$
terms
with fixed points of multiplicity
$\#\aut v$;
moreover, all the
$n/\#\aut v$
terms belonging to a given orbit give the same contribution to the sum.

\newsection{Perturbative expansion}

Let
$d\mu_a(M_a)$
be a collection of gaussian measures and let
$d\mu(M_.)=\prod_{a_\in\smallzet_k}d\mu_a(M_a)$.
Let us introduce the average
\beq
\la\quad\cdot\quad\ra=
{\int_{(\calh_N)^I}(\quad\cdot\quad)\,d\mu(M_.)
     \over\int_{(\calh_N)^I}\,\d\mu(M_.)},
\eeq
where
$\calh_N$
is the space of
$N\times N$
hermitian matrices.
The function
\beq
Z(\mu,c)=
\la\exp\tr\sum_{k=1}^K\sum_{\{a_.^k \} }
{c_k(a_1^k,\ldots,a^k_{l(k)})\over \#\aut v_k}
M_{a_1^k}\ldots M_{a_{l(k)}^k} \ra,
\eeq
where the $c_k$ are arbitrary coefficients,
can be expanded in a (formal) power series as
\bea
 Z(\mu,c)&=
\sum_{n_1,\ldots,n_K} \sum_{\{a_.^{k,r_.}\} }
&
\left(\prod_{k=1}^K{1\over n_k!\#\aut v_k^{n_k}}\right)\cdot \\
&&
c_k(a_1^k,\ldots,a_{l(k)}^k)^{n_k}\cdot
\la\prod_{k=1}^K\prod_{r_k=1}^{n_k}\tr
(M_{a_1^{k,r_k}}\ldots M_{a_{l(k)}^{k,r_k}})\ra. \nonumber
\eea
By Wick's theorem this decomposes in a sum over fat graphs.
We get
$\rho(\x)$
identical contributions for each isomorphism class
$[\x]$.
In view of the identity
\beq
\rho(\x)\cdot\#\aut\x=\prod_{k=1}^K n_k!\cdot(\#\aut v_k)^{n_k}
\eeq
which can be easily verified~\cite{tesi},
we finally get
\beq
Z(\mu,c)=\sum_{[\x]}{C(\x)\over\#\aut\x},
\label{stella}
\eeq
where
$\aut\x$
is the group of automorphisms of the graph
$\x$
with coloring $c:X_1\rightarrow\zet_k$ such as in~\rf{staruno},~\rf{stardue}
and
$C(\x)$
is computed according to the usual Feynman rules as a
product of contributions from the edges and vertices of the
graph
$\x$.
Eq.~\rf{stella}
has the form needed in order to represent the combinatorial
version of the integration of a volume form over the orbifold
${\cal S}_\gn^k$.

\newsection{Matrix integral}

We will now build up a matrix integral whose
perturbative expansion contains terms corresponding to colored
graphs verifying~\rf{staruno} and~\rf{stardue}.
The action
$S$
must satisfy the following conditions:

{\it i)}
terms of the form
$\tr(M_{a_1}\ldots M_{a_n})$
must come with a weight
$1/\#\aut v$
in order to implement~\rf{stella};

{\it ii)}
terms of the form
$\tr(M_{a_1}\ldots M_{a_n})$
with
$\sum_{i=1}^n a_i\not\equiv 0\;(\hbox{mod}\,k)$
should not appear in order to verify~\rf{staruno};

{\it iii)}
only propagators of the form
$\la M_aM_{-a}\ra$
must be different from zero in order to verify~\rf{stardue}.

\noindent
The three conditions are easily realized:

{\it i)}
{}From~\rf{dot}
we see that the weight
$1/\#\aut v$
naturally comes from terms of the form
$\tr(\sum c_a M_a)^n$;

{\it ii)}
terms with
$\sum_{j=1}^n a_j\not\equiv 0\;(\hbox{mod}\,k)$
can be canceled by choosing
$c_a=\exp({ia\phi\over R})$
and taking residues: we will then have a sum of terms of the form
\beq
\int_0^{2\pi R}{d\phi\over 2\pi R}
\left(\sum_a\exp({ia\phi\over R})M_a\right)^n,
\eeq
with $R$ a new positive real parameter.

{\it iii)}
in order to implement the third condition we will take in the
quadratic part only terms of the form
$\tr M_aM_{-a}$
and extend integration to the space of matrices
$M_a$
such that
$M_a=M_{-a}^\dagger$.

\noindent
More precisely, let us define the space of matrix trigonometric polynomials
\beq
\calt^+_{2k+1}=\{M(\phi):S^1_R\rightarrow gl(N)\; ;\;
M(\phi)=\sum_{a=-k}^k\exp({ia\phi\over R})M_a,\;
M_a=M_{-a}^\dagger,\; M(\phi)>0\}.
\eeq
(In the following we will restrict for notational convenience to the
$\zet_{2k+1}$-case).
Take
\beq
\cald M(\phi)=\prod_{a=-k}^k dM_a
\eeq
and define
\beq
Z_{2k+1}=
{\int_{\calt_{2k+1}^+}\cald M(\phi)\exp(-S[M(\phi)])
\over\int_{\calt_{2k+1}^+}\cald M(\phi)\exp(-{1\over 2}Q[M(\phi)])},
\eeq
where
$S$
can be chosen
{\it e.g.}
to be given by~\rf{pen} or~\rf{kon}
and
\beq
{Q\over 2}={1\over 2}\int_0^{2\pi R}{d\phi\over 2\pi R}\tr[M(\phi)]^2
=\tr\left({M_0^2\over 2}+\sum_{a=1}^k M_aM_{-a}\right),
\eeq
which gives
$\la(M_a)_{ij}(M_b)_{kl}\ra=\delta_{b,-a}\delta_{il}\delta_{jk}$
in the case~\rf{pen}
and the obvious generalization
$Q_\Lambda$
in the case~\rf{kon}.

\noindent
{\it Extension of the Penner model:}
Take
\bea
\label{pen}
S
&=& -Nt\int_0^{2\pi R}{d\phi\over 2\pi R}\tr[M(\phi)+\log(1-M(\phi))] \\
\nonumber
&=& {Q\over 2}+Nt\sum_{v\in V'}{1\over\#\aut v}
     \tr(M_{a_1(v)}\ldots M_{a_n(v)}),                    \\ \nonumber
\eea
where
$V=\cup_{n=3}^{+\infty}V_n$
and
\beq
V'_n=\{[(a_1,\ldots,a_n)]\in V_n:\sum_{j=1}^n a_j\equiv 0\; (\hbox{mod}\,k)\}.
\eeq
Repeating Penner's computation we finally get
\beq
\log Z_{2k+1}(N,t)=\sum_\gn\chi({\cal S}_\gn^{2k+1})\cdot
N^{2-2g}t^{2-2g-n},
\eeq
where $\chi({\cal S}_\gn^{2k+1})$ is now
the virtual Euler characteristic of the
space ${\cal S}_\gn^{2k+1}$.~\footnote{
The virtual Euler characteristic is the generalization to orbispaces
of the usual Euler characteristic, and it is obtained summing
the weight function
$(-1)^{\mbox{\scriptsize dim}\,c}/\#\aut c$
over all the orbicells $c$ of the orbispace
(in the case of ordinary spaces it would be just
$(-1)^{\mbox{\scriptsize dim}\,c}$).
}

\noindent
{\it Extension of the Kontsevich model:}
Take
\bea
\label{kon}
S
&=& -\int_0^{2\pi R}{d\phi\over 2\pi R}
     \tr\left(-{\Lambda\over 2}[M(\phi)]^2+
     {i\over 6}[M(\phi)]^3\right)       \\ \nonumber
&=& {Q_\Lambda\over 2}-{i\over 2}\sum_{v\in V_3'}
     {1\over\#\aut v} \tr(M_{a_1(v)}M_{a_2(v)}M_{a_3(v)}). \\ \nonumber
\eea
Repeating Kontsevich computations we get
\beq
\log Z_{2k+1}=\sum_{k_.}\la\tau_0^{k_0}\tau_1^{k_1}\ldots
\ra_{{\cal S}_\gn^{2k+1}}
     \prod_{i=0}^{+\infty}{t_1^{k_i}\over k_i!}
\eeq
where the subscript indicates that the Chern forms
$\tau_0,\tau_1,\ldots$
are now integrated over
${\cal S}_\gn^{2k+1}$
instead than over
$\mgn$.

\newsection{The kinetic term}

The extensions~\rf{pen} and~\rf{kon}
respectively of the Penner and Kontsevich model are trivial since
${\cal S}_\gn^{2k+1}$
is just a non-branched covering of
$\mgn$ made-up of $(2k+1)^{2g+n-1}$ sheets homeomorphic to ${\cal M}_\gn$.
This can also be seen directly by discrete Fourier transforming,
{\it i.e.}
by taking
$\hat M_j=\sum_{a=-k}^k\epsilon^{ja}M_a$
with
$\epsilon=\exp{2\pi i\over 2k+1}$,
$j=-k,\ldots,k$,
and noticing that
\beq
\sum_{j=-k}^k(\hat M_j)^n
=\sum_{j=-k}^k\left(\sum_{a=-k}^k\epsilon^{ja}M_a\right)^n
=\int_0^{2\pi R}{d\phi\over 2\pi R}
     \left(\sum_{a=-k}^k\exp({ia\phi\over R})M_a\right)^n.
\eeq
This means that each
$Z_{{\cal S}_\gn^{2k+1}}$
factorizes in a
$(2k+1)$-fold
product of
$Z_\mgn$.

However, we are now in a position to make a minimal extension to these
trivial models by adding to
$Q/2$
a kinetic term of the form
\beq
\label{kinetik}
{1\over 2}\int_0^{2\pi R}{d\phi\over 2\pi R}
     \tr[\dot M(\phi)]^2=\sum_{a=1}^k {a^2\over R^2}M_aM_{-a}.
\eeq
This gives the propagator
\beq
\la(M_a)_{ij}(M_b)_{kl}\ra=\delta_{a,-b}
{\delta_{il}\delta_{jk}\over 1+a^2/R^2},
\eeq
meaning that high values of
$a$
are disfavoured,
{\it i.e.},
cycles on the surface repels each other.
This gives a simple dynamics of cycles on surfaces.

The new partition function, which is now defined as an integral
over
${\cal S}_\gn^{2k+1,\comb}$
does no more project to a linear combination of integrals
over
${\cal S}_\gn^{2k+1}$,
since the product
$\prod_{x\in X}{1\over 1+a^2(x)/R^2}$
now depends
(as an almost everywhere constant function)
on the lengths
$p_i$
of the boundary components of the graphs.

The matrix models that are obtained by adding the term~\rf{kinetik}
to~\rf{pen} or to ~\rf{kon} are
representations of \( c=1 \) matter coupled to two-dimensional
quantum gravity with target space being a circle of radius \( R \).
This model has been studied by Gross and Klebanov~\cite{grosskleb},
using results of Marchesini and Onofri~\cite{marchesini}.
For \( R\rightarrow+\infty \) the non-singlet sector of the matrix
model decouples, since the non-singlet state energies are
logarithmically divergent in the double scaling limit.
The non-singlet degrees of freedom are identified with vortices on
the world sheet, which are dynamically irrelevant for large radius,
but condense at a critical value of the radius, giving rise to a
Kosterlitz-Thouless phase transition.
The complete free energy is given by a sum
\beq
F=F_{\mbox{\scriptsize s}}+F_{\mbox{\scriptsize ns}},
\eeq
where the non-singlet correction to the free energy has the simple
form
\beq
F_{\mbox{\scriptsize ns}}=
-\frac{1}{2\pi R}N^2 \exp\left(-2\pi R\,\delta(\mu))\right)
\eeq
with \( \delta(\mu))\sim|\log\mu| \)
in the continuum limit \( \mu\rightarrow+\infty \),
\( N\mu=\mbox{const} \).
In~\cite{grosskleb} $\mu$ plays the role of an UV-cutoff, and it has to
be identified with $k_0=k/R$ in our context. For $k_0\rightarrow+\infty$
we have thus $F_{\mbox{\scriptsize ns}}\sim N^2/Rk_0^{2\pi R}$.
Letting \( k\rightarrow+\infty \) with fixed $R$ the groups
$H_1(C,\zet_k)$ give the group $H_1(C,\zet)$.
If we instead let $k\rightarrow+\infty$, $R\rightarrow+\infty$ with
$k_0=\mbox{const}$ the groups $H_1(C,\zet_k)$ go over to
$H_1(C,\mbox{\bf R}_{k_0})$ with $\mbox{\bf R}_{k_0}=\mbox{\bf R}/k_0\zet$.
In our construction the parameter $R$ interpolates between the
topologically trivial situation $R=+\infty$ and a new topological
phase which has yet to be investigated.
We conjecture that the model at the critical point
$R=R_{\mbox{\scriptsize c}}=1$
should describe some non-trivial topological characteristics of the
space ${\cal S}_\gn$.

\begin{center}
{\bf Acknowledgments}
\end{center}
We thank E.~Arbarello, M.~Cornalba, R.~Dijkgraaf and C.~Procesi for
inspiring discussions, and E.~Witten for his comments about the
relation of our work with the model described in Ref.~\cite{witten}.
One of us (A.G.) is pleased to thank B.~Dubrovin for
long discussions had on the subject and for his warm support.

\end{document}